\documentclass{aa}

\usepackage{graphicx}
\usepackage{txfonts}
\usepackage{units}
\usepackage{hyperref}
\usepackage{subfigure}

\newcommand{\lsi}{LS~I~+61$^\circ$303}
\newcommand{\gammaray}{$\gamma$-ray}

\begin{document}

\title{
  Owens Valley Radio Observatory monitoring of \lsi{} completes three cycles of the super-orbital modulation
}
\titlerunning{
  OVRO monitoring of \lsi{} completes three cycles of the super-orbital modulation  
}

\author{F. Jaron\inst{1,2}
  \and
  S. Kiehlmann\inst{3}
  \and
  A. C. S. Readhead\inst{4}
}

\authorrunning{
  F. Jaron et al.
}

\institute{
  Technische Universität Wien,
  Wiedner Hauptstra\ss{}e 8-10,
  1040 Vienna,
  Austria, 
  \email{Frederic.Jaron@tuwien.ac.at}
  \and
  Max-Planck-Institut f\"ur Radioastronomie,
  Auf dem H\"ugel 69,
  53121 Bonn,
  Germany
  \and
  Institute of Astrophysics, Foundation for Research and Technology-Hellas, GR-71110 Heraklion, Greece, (2) Department of Physics, Univ. of Crete, GR-70013 Heraklion, Greece
  \and
  Owens Valley Radio Observatory, California Institute of Technology, Pasadena, CA 91125, USA 
}

\date{Received September 04, 2023; accepted January 18, 2024}

\abstract
{
  The high-mass X-ray binary \lsi{} is composed of a Be-type star and a compact object in an eccentric orbit. The emission from this source is variable and periodic across the electromagnetic spectrum, from radio to very high-energy gamma rays. The orbital period has been determined as $P_1 \approx 26.5$~d, and the source also features a super-orbital period with a value of $P_{\rm long} \approx 4.6$~years. Long-term monitoring of the binary by the Owens Valley Radio Observatory (OVRO) at 15~GHz has now completed 13.8~years, which corresponds to three full cycles of the super-orbital period. This is exactly one long-term cycle more than in the previous publication about OVRO observations of this source.
}
{
  Our aim is to investigate the presence and the stability of periodic signals in the radio data and to test if they are in agreement with previous results. This will contribute to the understanding of the physical processes behind the non-thermal emission from this source.
}
{
  We performed a timing analysis of the OVRO radio light curve and made use of the generalized Lomb-Scargle periodogram. We also combined the OVRO data with the full archive of previous radio observations and computed the discrete autocorrelation function.
}
{
  The most powerful features in the periodogram of the OVRO data are two peaks at $P_1 = \unit[26.49 \pm 0.05]{d}$ and $P_2 = \unit[26.93 \pm 0.05]{d}$, which are well separated from each other and clearly stand out above the very low noise level. The previously detected long-term period is still present in these new radio data, and our measurement is $P_{\rm long} = \unit[1698 \pm 196]{d}$. Dividing the OVRO data into three segments of equal length showed that the two periods, $P1$ and $P2$, are present in the periodogram of each of the consecutive long-term cycles. Our analysis of the full radio archive resulted in the detection of the same three periods, and the autocorrelation function showed a regular pattern, proving the continuity of the decades-spanning stability of the super-orbital modulation. In addition, we report a possible systematic modulation of the radio flux density with a timescale of approximately 40~years that has so far remained unnoticed.
}
{
  The physical model of a relativistic jet whose mass loading is modulated with the orbital period $P_1$ and is precessing with the slightly larger period $P_2$, giving rise to a beating with period $P_{\rm long}$, had previously been able to reproduce the radio and gigaelectron volt emission from this source. The ongoing presence and the stability of the periodic signals imply that this model is still the most plausible explanation for the physical processes at work in this source.
}


\maketitle

\nolinenumbers

\section{Introduction}

An X-ray binary is a stellar system composed of a normal star and a compact object, which can be either a neutron star or a black hole. Some X-ray binary systems are particularly bright in the gamma ray regime \citep{Dubus2013}, and where and how this high-energy emission is produced is still a debate. The high-mass X-ray binary \lsi{} is composed of a Be-type star and a compact object \citep{Casares2005}. The system is a source of emission across the electromagnetic spectrum, from radio to the very high-energy \gammaray{}s \citep[][and references therein]{Jaron2021}, and is also included in the search for neutrino emission \citep[][and references therein]{Abbasi2022}. 

The timing characteristics of its electromagnetic emission make \lsi{} special among the class of \gammaray{} emitting stellar binary systems. Radio emission was first detected from this source during a search for variable radio sources along the galactic plane \citep{Gregory1978}. Since then, it has been the target of repeated radio monitoring, which has revealed that it is not only variable but also periodic on different timescales. The most accurate measurement of the orbital period was obtained by \citet{Gregory2002}, who analyzed radio data from a long-term monitoring program of the Green Bank Interferometer (GBI) at~2 and~8~GHz and reported a value of $P_1 = \unit[26.4960 \pm 0.0028]{d}$. In the same article, \citet{Gregory2002} also firmly established that the radio emission from this source is subject to a super-orbital long-term modulation with a period of $P_{\rm long} = 1667 \pm 8$~d. This super-orbital modulation is present across the spectrum until the teraelectron volt regime, with a systematic phase relationship between wavelengths \citep{Jaron2021}. In particular, the large database in the radio has proven that this long-term modulation has remained stable over decades of observations \citep{Massi2016}. 

Concerning the eccentricity of the orbit, analyses of optical data from \lsi{} have produced mixed results. \citet{Casares2005} obtained a value of $e = 0.72 \pm 0.15$ using absorption lines that were least contaminated by emission lines from the Be circumstellar disk. \citet{Aragona2009} analyzed the He~I~$\lambda$6678 absorption line, which resulted in $e = 0.537 \pm 0.034$. However, this absorption line was slightly affected by blue and red wings in emission. More recently, \citet{Kravtsov2020} reported a value of only $e \approx 0.1$, which they determined by model fitting the optical linear polarization of the source. However, their reported value is at odds with other observations. This outcome can be understood considering the approach used by \citet{Kravtsov2020}, as they used a simple model of variations induced by the orbital motion of the compact star with respect to the decretion disk of the Be star, and they did not include the other disk in their model, i.e., the disk around the compact object. For these reasons, we assumed a high eccentricity of the orbit for the interpretation of our results and considerations about the physical processes at work in this source.

Evidence for periodic accretion and ejection along the orbit of \lsi{} has been given in \citet{Massi2020}, and the associated X-ray luminosity is related to its photon index, as established for accreting black holes at all masses \citep{Massi2017}. Evidence for the presence of a jet in \lsi{} comes from direct very long baseline interferometry (VLBI) observations \citep{Massi2012, Wu2018}. Indirect evidence has been obtained by analyzing the radio spectrum. A six-year archive of radio observations was analyzed by \citet{Massi2009}, revealing the characteristics of radio jet emission, namely, a radio peak with a flat spectrum followed by an optically thin radio outburst, which is typically observed in microquasars (i.e., in accreting black holes or neutron stars with a low magnetic field) \citep{Mirabel1999}. This is also in agreement with the expected two-peaked \citet{Bondi1944} accretion rate profile along the eccentric orbit of this source, as outlined by several authors \citep{Taylor1992, Marti1995, Bosch-Ramon2006, Romero2007}. A physical model of a self-absorbed jet, which through precession gives rise to periodic changes in the Doppler boosting of the intrinsic emission reproduces 37~years of radio observations and its spectral index \citep{Massi2014}. The inclusion of external inverse Compton scattering and synchrotron self-Compton into that model reproduces several years of simultaneous radio and \textit{Fermi}-LAT $\gamma$-ray observations and their timing characteristics \citep{Jaron2016}. This model also explains why the radio emission peaks only once per orbit (i.e., at apastron), the reason being that at periastron the relativistic electrons of the jet suffer catastrophic inverse Compton losses in the strong UV photon field in the proximity of the Be star. Finally, the presence of the afore mentioned super-orbital modulation at all observed wavelengths and a systematic phase offset between wavelengths finds a straightforward explanation in a precessing jet in which the higher energy emission is produced upstream from the lower energy emission \citep{Jaron2021}. Nevertheless, there is also a pulsar scenario being discussed for \lsi{} as a generalization of the phenomena observed in the pulsar binary PSR B1259-63 to the larger class of gamma ray binaries \citep{Dubus2013}.

The possibility that the physical processes behind the non-thermal emission from \lsi{} are powered by the spin-down of a millisecond pulsar have been discussed since soon after the discovery of the source \citep{Maraschi1981}. However, despite dedicated searches, pulses were never detected from this source \citep{Canellas2012}. The only signal that pointed to the presence of a pulsar in the system was a short X-ray burst that was interpreted as a "magnetar-like event" by \citet{Torres2012}. This signal was observed from a region on the sky that contains several sources other than \lsi{}, putting into question the association with that particular source. More recent observations with the Five-hundred-meter Aperture Spherical radio Telescope (FAST) have resulted in the detection of 42 radio pulses from the direction of \lsi{}, with a pulse period of $P = 269.15508 \pm 0.00016$~ms \citep{Weng2022}.
The field of view of these observations is even larger than that of \citet{Torres2012}, so the possibility of the pulses originating from another source cannot be ruled out. Furthermore, the pulses were detected during an orbital phase of $\Phi = 0.59$ with an exposure time of three hours (see Table~1 in \citealt{Weng2022}). Two subsequent observations listed in their Table~1 made with the same telescope during similar orbital phases ($\Phi = 0.58, 0.62$) and with similar exposure times (3 and 2~h, respectively) did not result in the detection of any pulses. This means that, so far, there is not any proof of a dependency of the occurrence of pulses on the orbital period of the system. Finally, \citet{Weng2022} did not detect any Doppler shift of the pulses, which would indicate that they do not originate from a pulsar in a binary system at all. In conclusion, there is not enough evidence to make a firm association between the observed pulses and the astrophysical object~\lsi{}. In their supplementary Fig.~1, \citet{Weng2022} reported a spin-down of $\dot{P} = (4.2 \pm 1.2)\,10^{-10}$ that should be regarded as tentative because it does not result from direct timing, as also pointed out by \citet{Suvorov2022}. Assuming this value would imply a spin-down power of $\dot{E} = (8.5 \pm 2.4)\,10^{38}$\,erg/s, which is indeed a relatively large value, but it is not reliable because of the questionable measurement of $\dot{P}$ itself. More observations would be needed to measure the spin-down with a higher accuracy and to probably disentangle it from the orbital motion if the detected pulsar is indeed part of a binary system. 

Even if \lsi{} contained a pulsar, this would not rule out the possibility of a jet. Pulsar wind nebulae can have flat radio spectra and jets (see, e.g., \citealt{Slane2017} for a review). Extended X-ray emission close to the $\gamma$-ray-loud binary system LS~5039 has been interpreted as a pulsar wind nebula by \citet{Durant2011}. For \lsi{}, however, such an observation has not been reported anywhere as of yet. Only the extended structure in the aforementioned VLBI images \citep{Massi2012} has previously been interpreted as a "cometary tail" by \citet{Dhawan2006}. The reported values of $P$ and $\dot{P}$ by \citet{Weng2022} implying a surface magnetic field on the order of $10^{14}$\,G would certainly inhibit the formation of a radio jet (\citealt{Massi2008} determined an upper limit of $10^7-10^8$\,G for the formation of a radio jet from a neutron star). Such a large magnetic field is therefore in contradiction to the observation of a radio jet that has a self-absorbed spectrum \citep{Massi2009, Zimmermann2015}, which is the spectrum that is typical of microquasars. Observations of radio emission from the millisecond pulsar binary PSR~J1023+0038, which has a magnetic field of $\sim 10^8$\,G \citep{Deller2012}, has recently been provided by \citet{Baglio2023} and interpreted as originating from a radio jet.

The precise timing characteristics of \lsi{} make this source a unique laboratory to study the physical processes behind the non-thermal emission. In particular, the presence of the super-orbital modulation sets this source apart from other objects in the class of \gammaray{} emitting X-ray binaries. In this article, we present new radio observations of \lsi{} carried out with the Owens Valley Radio Observatory (OVRO) at 15~GHz \citep{Richards2011}. With a baseline of 13.8~years, these observations have now completed three cycles of the 4.6-year super-orbital modulation. This represents exactly one super-orbital cycle more of data than was available at the time of the previous publication about OVRO monitoring of this source presented in \citet{Jaron2018}. The first aim of our analysis in this work is to investigate the presence of periodic signals in the updated OVRO data. The second aim is to test the stability of periodic signals, and the long-term modulation in particular, by combining the new OVRO observations with the radio archive from \citet{Massi2016}. 

The paper is structured as follows. In Sect.~\ref{sec:data}, we present the data sets. In Sect.~\ref{sec:methods}, we describe the methods that we used for data analysis. We present our results in Sect.~\ref{sec:results} and discuss them in Sect.~\ref{sec:discussion}. We give our conclusions in Sect.~\ref{sec:conclusions}.

\section{The data} \label{sec:data}

\subsection{OVRO monitoring}

The OVRO regularly observes \lsi{} at 15~GHz as part of a  monitoring program \citep{Richards2011}. The radio light curve that we used for our analysis covers the time span modified Julian day (MJD)~54908-59961 (2009 March 18--2023 January 17). This represents 13.8~years of data, which is three cycles of the long-term modulation of \lsi{}. The exact super-orbital phase range of the data is $\Theta = 6.92 - 9.96$ (as defined in Sect.~\ref{sec:periods}). A description of the data calibration can be found in Sect.~2.1 of \citet{Jaron2018}, and for further details, we refer to \citet{Richards2011}.

\begin{figure*}
  \centering
  \includegraphics{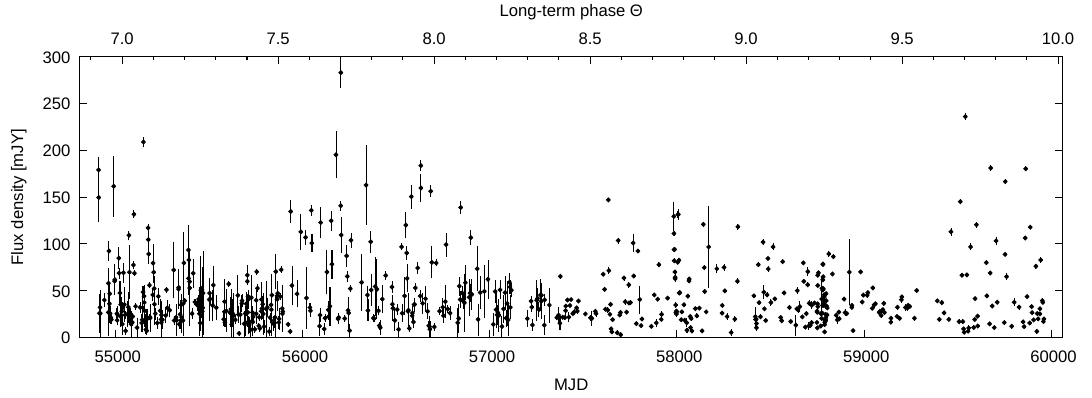}
  \caption{
    Radio light curve from OVRO monitoring of \lsi{} at 15\,GHz. The flux density is plotted against the time of observation given in MJD in the lower axis and in cycles of long-term modulation in the upper axis. Only data points above $1\sigma$ are plotted here and were included in the analysis for this work. 
  }
  \label{fig:ovro-lc}
\end{figure*}

Figure~\ref{fig:ovro-lc} shows the OVRO flux density plotted against time, given in MJD in the lower $x$-axis. The upper $x$-axis shows the long-term phase $\Theta$. For the analysis of this present work, we selected data above $1\sigma$, meaning that the value of their flux density is greater than their uncertainty. Flux variability and a long-term modulation pattern can be seen in the light curve. A quantitative analysis of the flux variability is the subject of our investigation.

\subsection{Archival data}

\begin{figure*}
  \includegraphics{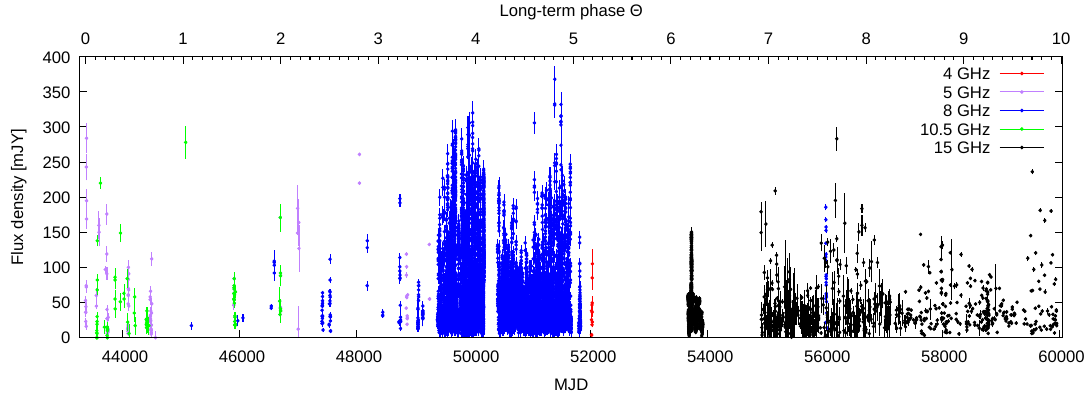}
  \caption{
    Concatenation of all available radio data for \lsi{} between 4 and 15~GHz. The flux density is plotted against time, and the observed radio frequency is indicated by color.
  }
  \label{fig:radioall-lc}
\end{figure*}

Since its first radio detection on 1977 August 11 \citep{Gregory1978}, \lsi{} has been the subject of repeated radio observations. \citet{Massi2016} compiled all available 4 to 15~GHz radio data from this source for a detailed timing analysis. Here, we extend this database with the new OVRO observations described above. The resulting long-term light curve is shown in Fig.~\ref{fig:radioall-lc}. These observations were carried out at different radio frequencies. For a better visual display in the figure, we rounded the frequencies to half gigahertz steps and color coded them as indicated by the legend in the plot. These archival radio data span the time range from MJD~43367.480 until MJD~59961.318, which is a total of 45.5~years. The sampling of the data is however impacted by irregularity from 
over the years and different observing programs, as is immediately visible in the plot.

\section{Methods} \label{sec:methods}

\subsection{Lomb-Scargle periodogram}

The data we used for our analysis are not regularly sampled, which is a typical situation in radio astronomy. The average sampling rate of the OVRO data set, however, is 6.7~days, which is well below the Nyquist limit for the range of the shortest periods that we are interested in ($P \sim 26-27$~d). More details about the sampling of the OVRO data can be found in Appendix~\ref{app:sampling}.

For our timing analysis, we made use of the implementation of the Lomb-Scargle periodogram \citep{Lomb1976, Scargle1982} in the \texttt{astropy} Python package \citep{astropy2013, astropy2018, astropy2022}. Details about the estimation of the false alarm probability of features in the periodograms can be found in Appendix~\ref{app:fap}. In order to determine the center period and uncertainty of peaks, we carried out functional fits to the periodogram, as described in Appendix~\ref{app:fit}. A potentially more robust method of computing the Lomb-Scargle periodogram has been introduced by \citet{Zechmeister2009} and implemented in Python by \citet{Czesla2019}. We also applied their method to our data for comparison, and we found that the differences between the results obtained with their method and those obtained with the \texttt{astropy} package are marginal.

\subsection{Discrete cross-correlation function}

A powerful tool to quantify the self-similarity of data is the auto-correlation function, which we realized as the discrete cross-correlation function (DFC) of the data with themselves. The irregular sampling of the data also had to be accounted for in this context. We computed the DCF with the help of a \texttt{Python} implementation of the \citet{Edelson1988} method by \citet{Robertson2015}.

\subsection{Intrinsic periods and epoch folding} \label{sec:periods}

The orbital phase of the binary system \lsi{} is defined as
\begin{equation} \label{eq:Phi}
  \Phi = \frac{t - t_0}{P_1} - \mathrm{int}\left(\frac{t - t_0}{P_1}\right),
\end{equation}
where $t$ is the time of observation, $t_0 = 43366.275$~MJD is the epoch of first radio detection of the source, $P_1$ is the orbital period, and $\mathrm{int}(x)$ takes the integer part of $x$, so $\Phi$ is a real number between zero and one. Periastron has been estimated to occur at the orbital phase $\Phi = 0.23 \pm 0.02$ \citep{Casares2005}.

\citet{Gregory2002} estimated the value of the long-term modulation with high accuracy as $P_{\rm long} = \unit[1667 \pm 8]{d}$. The long-term phase is obtained by replacing $P_1$ with $P_{\rm long}$ in Eq.~(\ref{eq:Phi}). It is occasionally useful to keep the integer part. The phases then represent the number of cycles elapsed since MJD~43366.275. Phases of other periodic signals from \lsi{} are defined in an analogous way by inserting the corresponding period value.

In order to investigate the shape of the periodic signals, we made use of the epoch folding technique. For this purpose, we plotted the data against the phase of the corresponding period. A variant of this is to put the data into phase-bins for averaging first.

\section{Results} \label{sec:results}

\subsection{Spectral analysis} \label{sec:results-spec}

\begin{figure*}
  \includegraphics{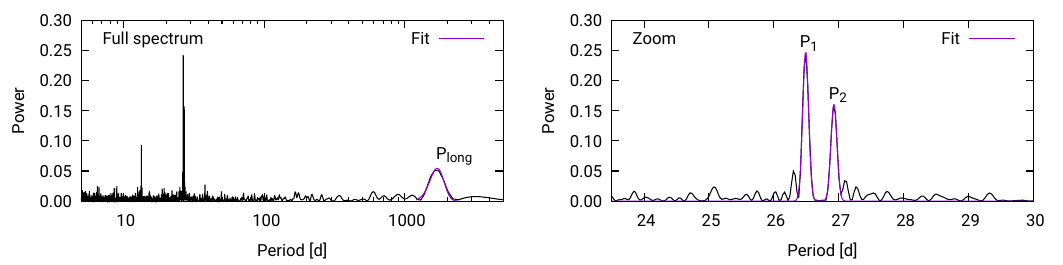}
  \caption{
    Lomb-Scargle periodogram of the OVRO data (shown in Fig.~\ref{fig:ovro-lc}). Left: Full spectrum showing a peak at the long-term period, a feature at the orbital period, and a harmonic of it at $\sim 13$\,d. Right: Zoom into the orbital range revealing a two-peak feature.
  }
  \label{fig:ovro-sc}
\end{figure*}

The Lomb-Scargle periodogram of the OVRO data (plotted in Fig.~\ref{fig:ovro-lc}) is shown in Fig.~\ref{fig:ovro-sc}. The left panel shows the full spectrum. The strongest feature is in the position of the orbital period ($\sim 26.5$~days), and a harmonic of this is also present at half the period. And finally there is also a peak in the position of the long-term period ($\sim 1667$~days). The right panel of Fig.~\ref{fig:ovro-sc} shows a zoom into the region around the orbital period. Two very distinct peaks are present that are well detached from each other and clearly stand out above the very low noise level. The presence of these two peaks is in agreement with previous findings, as is explained in Sect.~\ref{sec:interpretation}.

In order to quantify the positions of the two peaks, we fit the periodogram in the range of interest with the superposition of two Gaussian functions, as described in Appendix~\ref{app:fit}. The estimates of the two periods are $P_1 = 26.49 \pm 0.05$~d and $P_2 = 26.93 \pm 0.05$~d. Details of the resulting fit parameters are given in the upper part of Table~\ref{tab:fitscovro}, and the fitted function is plotted as the solid purple curve in Fig.~\ref{fig:ovro-sc}. The agreement between the data and the fit is remarkable.

We performed a fit of a single Gaussian to the peak at the long-term period (visible in the left panel of Fig.~\ref{fig:ovro-sc}) and obtained the results shown in the lower part of Table~\ref{tab:fitscovro}. Our estimate of the long-term period is $P_{\rm long} = 1698 \pm 196~\mathrm{d}$. Again, the fitted function is plotted as the purple solid curve in the figure.

To determine the significance of peaks in the Lomb-Scargle periodogram, we made use of the functionality in the \texttt{astropy} package to compute the false alarm probability~$p$ of peak powers. Figure~\ref{fig:fap} shows the false alarm probability plotted against peak power. The peak powers of the periods $P_1$, $P_2$, and $P_{\rm long}$ are marked by the vertical dashed lines. For the data shown in Fig.~\ref{fig:fap}, the \citet{Baluev2008} method has been used, which is the default option of the \texttt{astropy} package. The dotted horizontal line is at the height of a probability of $p = 0.1$, which is conventionally the upper limit for a periodogram feature to be considered significant \citep{LinnelNemec1985, Baluev2008}. All three peaks have $p$-values well below this limit, which means that they are significant. The probabilities for $P_1$ and $P_2$ are numerically zero. For $P_{\rm long}$, we obtained $p = 10^{-4}$, which is still very significant. Besides the \citet{Baluev2008} method, three other options are available in \texttt{astropy}, all of which gave consistent results concerning the significance of the three periods.

\subsection{Signal profiles} \label{sec:results-fold}

\begin{figure}
  \includegraphics{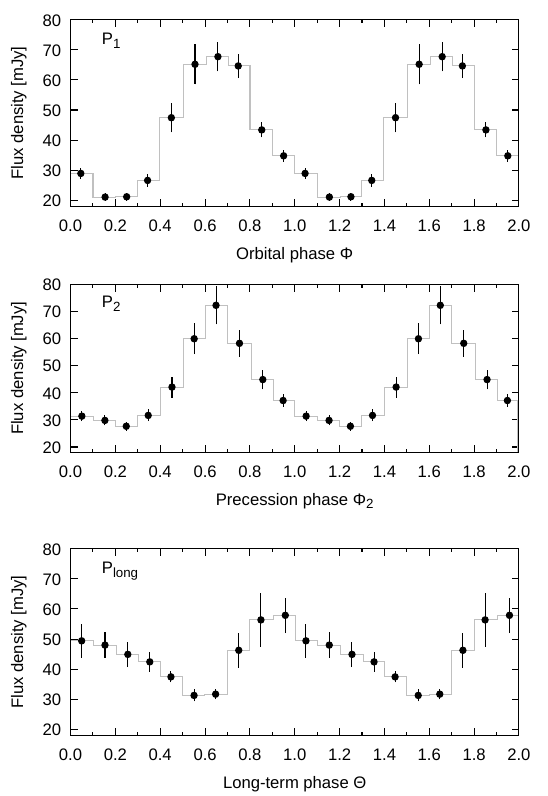}
  \caption{
    Variability profiles of the periodic signals in the OVRO data (shown in Fig.~\ref{fig:ovro-lc}).
    Top:~Data folded over $P_1$, showing the orbital profile.
    Middle:~Data folded over $P_2$, showing the precession profile.
    Bottom:~Data folded over $P_{\rm long}$, showing the profile of the long-term modulation.
  }
  \label{fig:ovro-fold}
\end{figure}

To investigate the profiles of the signals that are periodic with $P_1$, $P_2$, and $P_{\rm long}$, we folded the OVRO data over these periods. We put the data into ten phase-bins and averaged the flux values in each bin. The folded light curves are shown in Fig.~\ref{fig:ovro-fold}, where the error bars represent the $1\sigma$ standard deviation of the distribution in each phase-bin. The phase of the $x$-axes range from zero to two, with values greater than one repeated from the unit interval. The top panel of the figure shows the data folded over the orbital period~$P_1$. In the middle panel, the data are folded over the precession period $P_2$, and the bottom panel shows the data folded over the long-term period $P_{\rm long}$. All three profiles show a distinct one-peaked pattern. The orbital profile peaks at $\Phi \approx 0.7$, the precession profile peaks at $\Phi_2 \approx 0.7$, and the long-term profile peaks at $\Theta \approx 1.0$. This is in agreement with previous findings reported in the literature \citep{Jaron2018} and is an indication of the stability of these patterns. In particular, we find it worth noting that the peak of the orbital profile is still at apastron, with periastron occurring at $\Phi = 0.23$ \citep{Casares2005}. This is not only in agreement with previous observations but also with the physical modeling of \citet{Massi2014} and \citet{Jaron2016}. Folding the data over an arbitrary period does not result in any distinct variability profile, as is shown in Appendix~\ref{app:foldrand}.

\subsection{Time-resolved analysis of the OVRO data}

\begin{figure}
    \centering
    \includegraphics{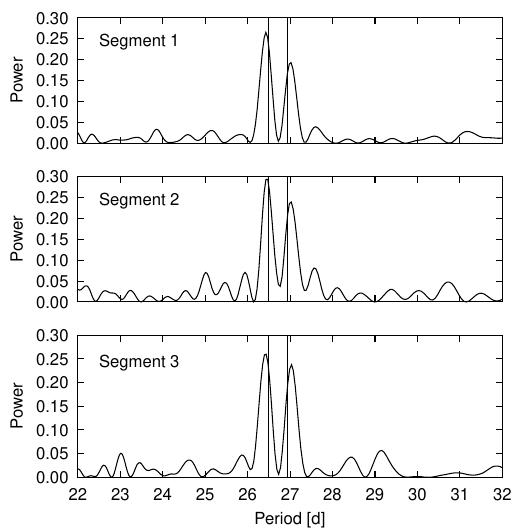}
    \caption{
        Lomb-Scargle periodograms of the OVRO data (shown in Fig.~\ref{fig:ovro-lc}) divided into three individual segments. The two peaks, which are found in the full data set, are also present in each of the segments. The vertical lines indicate the positions of $P_1$ and $P_2$ in the full OVRO data set.
    }
    \label{fig:LS3seg}
\end{figure}

The OVRO monitoring of \lsi{} has now completed three full cycles of the long-term period of the source. In order to analyze the spectral characteristics of each of the three long-term cycles, we divided the full OVRO data set into three segments of equal length. Each segment is 1684.2~d long. We performed generalized Lomb-Scargle timing analysis on each one of these segments. The resulting periodograms are shown in Fig.~\ref{fig:LS3seg}. In each of the periodograms, the two-peaked profile of $P_1$ and $P_2$ is clearly present, and the period values are in agreement with the ones found in the analysis of the entire OVRO data, as indicated by the two vertical lines, which correspond to the values reported in Sect.~\ref{sec:results-spec}. No significant deviation from these values was detected.

\subsection{Stability of the long-term modulation}

In this section, we investigate the stability of the long-term modulation of the radio emission from \lsi{} by analyzing the long-term radio light curve shown in Fig.~\ref{fig:radioall-lc}. This light curve is the combination of the new OVRO data with the entire radio archive from \citet{Massi2016}.

\begin{figure}
  \includegraphics{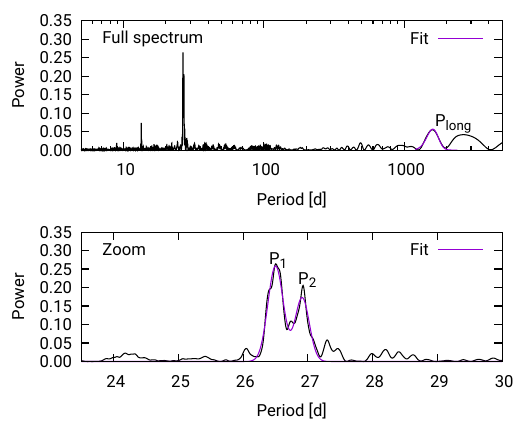}
  \caption{
    Lomb-Scargle periodogram of the entire light curve shown in Fig.~\ref{fig:radioall-lc}. The upper panel shows the entire spectrum, with a prominent feature at the position of the orbital period (and a harmonic at $\sim 13$\,d) and a feature at the position of the long-term period and a subharmonic of it. The lower panel shows a zoom into the region around the orbital period, revealing a two-peaked feature.
  }
  \label{fig:radioall-sc}
\end{figure}

We computed the Lomb-Scargle periodogram of the full data set with the same method that we applied to the OVRO data alone. The result is shown in Fig.~\ref{fig:radioall-sc}, where the upper panel shows the full spectrum and the lower panel shows a zoom into the region around the orbital period. The result is very similar to the periodogram of the OVRO data alone (c.f.~Fig.~\ref{fig:ovro-sc}) except that the subharmonic of $P_{\rm long}$ is more pronounced here, gaining almost the same power as the long-term period itself. In the lower panel, the zoom into the orbital period range shows a double-peaked profile. We fit this profile with the superposition of two Gaussian functions, which is analogous to what we did with the OVRO data above. The resulting periods are $P_1 = 26.50 \pm 0.12$~d and $P_2 = 26.91 \pm 0.11$~d. The fitted function appears as the solid purple curve in the figure. We performed a fit of a single Gaussian to the peak at the position of the long-term modulation and obtained $P_{\rm long} = 1601 \pm 152$~d. The detailed results of the fit parameters are listed in Table~\ref{tab:fitscall}.

\begin{figure}
  \includegraphics{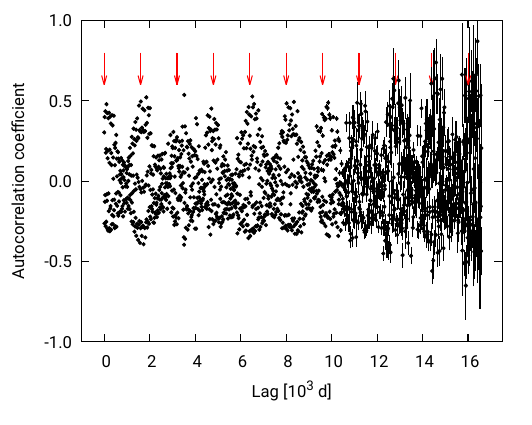}
  \caption{Autocorrelation function of the entire radio light curve (shown in Fig.~\ref{fig:radioall-lc}). The autocorrelation coefficient is plotted against the lag. Distinct peaks occur at integer multiples of the long-term period, marked by the red arrows.
  }
  \label{fig:radioall-dcf}
\end{figure}

Following up on the analysis by \citet{Massi2016}, we computed the discrete autocorrelation function of the long-term radio data. The result is shown in Fig.~\ref{fig:radioall-dcf}, where the autocorrelation coefficient is plotted against the time lag. The plot shows an oscillatory pattern with peaks at integer multiples of the long-term period, as highlighted by the red arrows. This shows that the radio data become self-similar when they are shifted against themselves by $n\cdot P_{\rm long}$, where $n$ is an integer number and $P_{\rm long} = 1601$\,d. With this plot, we confirmed the result previously obtained by \citet{Massi2016} in the left panel of their Fig.~5, where they showed that the long-term modulation had remained stable for eight full cycles. Our result now shows that this trend is ongoing and that the long-term modulation has by now remained stable for ten full cycles. 

\section{Discussion} \label{sec:discussion}

The values of the periodic features found by the timing analysis presented here are in very good agreement with previously published results. In particular, the periodogram of the third segment of the OVRO data (i.e., the new data only) shows that the two periods, $P_1$ and $P_2$, are still present in the most recent radio data, and that their values have remained stable (bottom panel of Fig.~\ref{fig:LS3seg}). In this section, we first discuss which physical processes are the possible mechanisms behind these periodic signals. We then show how these periodicities are mathematically connected through interference. And finally, we also briefly discuss properties of the full radio data set on even longer timescales.

\subsection{Physical processes behind the periodic features} \label{sec:interpretation}

The most precise measurement of the orbital period of the binary system~\lsi{} is still the value $P_1 = 26.4960 \pm 0.0028$~d determined by \citet{Gregory2002}, who applied Bayesian hypothesis testing to the radio data from the GBI monitoring. There are two possible reasons why his analysis did not also result in the detection of the nearby period $P_2 \approx 26.9$~d, as the timing analysis by \citet{Massi2013} of the same data set did. First of all, \citet{Gregory2002} did not test for the hypothesis of three periodic features (i.e., $P_1$, $P_2$, and $P_{\rm long}$).\footnote{We note that in his publication, \citet{Gregory2002} used the symbol $P_2$  for the long-term period, which we denote as $P_{\rm long}$.} Secondly, as shown by \citet{Massi2013}, the periodic feature $P_2$ has a lower power in the periodogram of the GBI data set than $P_1$, so if one only tests for one period in that range, the analysis will detect the stronger~$P_1$ and not the weaker~$P_2$.

\citet{Massi2013} interpreted the periodic feature at $P_2 = 26.92 \pm 0.07$~d as the precession period of a relativistic jet. The reason for this conclusion was that \citet{Massi2012} had previously analyzed a sequence of VLBA observations of \lsi{} and by applying the method of phase-referenced astrometry, they showed that the core component traces an ellipse and that it takes this ellipse longer than one orbital cycle to return to its initial position. This result was later confirmed by \citet{Wu2018}, who revisited the source with VLBI astrometry and showed that the core still traces an ellipse that overlaps with the previous observations by \citet{Massi2012} after correcting for the proper motion of the source. By analyzing the combined data set, \citet{Wu2018} determined the precession period of the core component as $P_{\rm precession} = 26.926 \pm 0.005$~d. This value is in remarkable agreement with the values from the timing analysis of different data sets at multiple wavelengths \citep{Massi2013, Jaron2014, Massi2016, DAi2016, Jaron2018}, all of which detected the double-peaked profile of $P_1$ and $P_2$ in the power spectra. A physical model of a self-absorbed relativistic jet precessing with a period $P_2$ and periodically refilled with a population of relativistic electrons with period $P_1$ is able to reproduce decades of radio observations and their spectral characteristics, as shown by \citet{Massi2014}. The physical process responsible for the modulation of the emission with $P_2$ is periodic changes in the Doppler boosting of the intrinsic emission by a jet that continuously and periodically changes the angle with respect to the line of sight of an observer on Earth. With the work presented in this article, we confirm that the precession signal is still active in the most recent radio emission from this source (as shown, in particular, in the bottom panel of Fig.~\ref{fig:LS3seg}).

In the periodograms presented in this article (Fig.~\ref{fig:ovro-sc} for the OVRO monitoring and Fig.~\ref{fig:radioall-sc} for the entire radio data set), the most powerful features are found at the positions of $P_1$ and $P_2$, as discussed above. The third prominent feature is a peak at the position of the known long-term period $P_{\rm long}$. The value of this period was determined with the highest precision by \citet{Gregory2002} as $P_{\rm long} = 1667 \pm 8$~d through Bayesian analysis of the GBI data. The timing analysis of our work here gives results that are, within their uncertainties, in agreement with this measurement. In addition, we repeated the autocorrelation analysis first carried out by \citet{Massi2016} on the updated data set of radio observations since 1977, the result of which is presented in Fig.~\ref{fig:radioall-dcf}. This result confirms the previously published result by \citet{Massi2016}, and it shows that the long-term modulation is still active in \lsi{} and is still stable.

\subsection{Periodic variability in the Be star disk}

The non-degenerate component in the binary system \lsi{} is a Be-type star \citep{Hutchings1981}. These stars have a high angular momentum, leading to an equatorial outflow in the form of a decretion disk, which is the source of emission lines seen in their optical spectra (see \citealt{Rivinius2013} for a review). The definition of the Be phenomenon already includes the possibility that the presence of emission lines can be variable, which implies that the disk itself can be subject to variability over time. Physical processes connected to precisely periodic Be star variability have been observed to occur on timescales of 0.5 to 2 days. To the best of our knowledge, there are not any (quasi-)periodic processes in Be star disks that occur on the timescales on the order of the period $P_2$, that is, a few tens of days. For this reason,
the possibility that $P_2$ itself may correspond to any physical process in the Be star disk is ruled out. There are, however, two processes that can occur on timescales on the order of a few years, which would be compatible with the long-term modulation $P_{\rm long}$ of \lsi{} and is why these processes deserve to be mentioned here. The first process is variability of the size of the Be star disk. The second is a one-armed density wave. Although both types of variability are often observed in Be X-ray binaries, neither of these has ever been observed to occur strictly periodically, especially not over a time span of several decades, which is how the long-term modulation in \lsi{} occurs.

In X-ray binaries, the decretion disk around a star displaying the Be phenomenon is not the only possible source of line emission. \citet[][and references therein]{Fender2009} presented observational evidence that the accretion disk around the compact object can be a source of strong emission lines. In their Fig.~1, \citet{Fender2009} present optical spectra of the black hole X-ray binary GX~339-4, which they state is almost certainly uncontaminated by any companion star. The spectra shown there correspond to different X-ray states of the source (faint, bright-hard, and bright-soft). While emission lines can be seen in all three presented spectra, they are the strongest in the faint state. In all three states, the strongest feature is the H$\alpha$ line, but there are also several He lines present. Since \lsi{} remains in a low-hard X-ray state, on the boundary to the quiescent state \citep{Massi2017, Massi2020}, it is very possible that in this system the accretion disk is also a significant source of line emission. That the optical emission lines indeed originate from a rotating flow is shown in Fig.~7 in \citet{Fender2009}, where it is highlighted that most of the observed emission lines have a double-peaked profile, which is discussed in more detail in their Sect.~3. In their conclusion, the authors state that "low-luminosity accreting sources should be clearly identifiable in, for example, H$\alpha$ surveys by their large [equivalent widths]." These observational facts should be kept in mind when interpreting the presence and the timing characteristics of any emission lines from \lsi{}. 

If the size of a Be star disk changes, then this manifests itself in a variability of the photometric light curve of the optical emission. In this context, a well-studied system is the Be X-ray binary A0538-66. \citet{Rajoelimanana2017} reported on how quasi-periodic variability in the optical emission from this source, especially pronounced in the V-band data shown in the upper panel of their Fig.~1, is related to variation in the size of the Be star disk. Concerning our target of interest, \lsi{}, such an optical light curve has never been reported anywhere. Furthermore, \citet{Rajoelimanana2017} show periodigrams (their Fig.~3) that reveal that there is only one peak at the orbital period. There is no two-peaked profile, which is unlike the periodograms we report (Figs~\ref{fig:ovro-sc}, \ref{fig:LS3seg}, and \ref{fig:radioall-sc}) and those that have been reported before, as explained in the previous subsection. The absence of a second peak in A0538-66 thus demonstrates that a quasi-periodic modulation of the Be star disk in the form of a gradual buildup and decay (as suggested for \lsi{} by \citealt{Chernyakova2012}) does not result in a beating with the orbital period. Another probe of the Be disk size is the equivalent width (EW) of the H$\alpha$ emission line. However, remembering that in X-ray binaries the accretion disk can also be a source of H$\alpha$ and other line emission, detections that EW(H$\alpha$) is modulated with $P_{\rm long}$ in \lsi{} \citep{Zamanov2013, Paredes-Fortuny2015} are not proof of that variability being related to periodic changes in the Be star disk. The interpretation of EW(H$\alpha$) is further complicated by the fact that both the accretion and the jet contribute their variable optical continuum to its normalization (as been pointed out in \citealt{Jaron2021}). In any case, a strictly periodic variability of a Be star disk size has never been observed, so it is an unlikely physical process behind the periodic long-term modulation of \lsi{}. 

Cyclic variation in a Be star disk in the form of a one-armed density wave manifests itself as modulation of the V/R (violet to red) ratio of optical emission lines. As \citet{Massi2016} already pointed out, a well-studied example of this type of variability is the Be binary sytem $\zeta$~Tau, as examined in detail by \citet{Stefl2009}. The database available for this system spans a century and shows that the modulation of the V/R ratio is quasi-periodic at times but can also be completely absent for a few decades. Concerning \lsi{}, a periodic modulation of the V/R ratio of any emission line has never been reported despite the searches performed by \citet{Zamanov1999, Zamanov2013}. The periodic behavior of the long-term modulation of \lsi{} makes it very unlikely that there is any connection to this type of Be star disk variability.

The period $P_2 = 26.926 \pm 0.005$\,d has been explicitly measured for the precession period of the VLBI core component by \citet{Wu2018}. This is why in the following section we explain in detail how the strictly periodic long-term modulation of \lsi{} fits into the scenario of a beating between $P_1$ and $P_2$.

\subsection{The long-term period as a beating between orbit and precession} \label{sec:beating}

\citet{Massi2013} were the first authors who reported the presence of $P_1$ and $P_2$ in the radio emission from \lsi{} and who interpreted the long-term period $P_{\rm long}$ as the result of the interference between the orbital period~$P_1$ and the precession period~$P_2$ in the form of a beating. In the following, we provide a short review of how the periods found in the power spectra, here and in previous publications, fit into the mathematical concept of a beating and how this corresponds to the characteristics of the radio emission from \lsi{}. 

\begin{figure*}
  \includegraphics{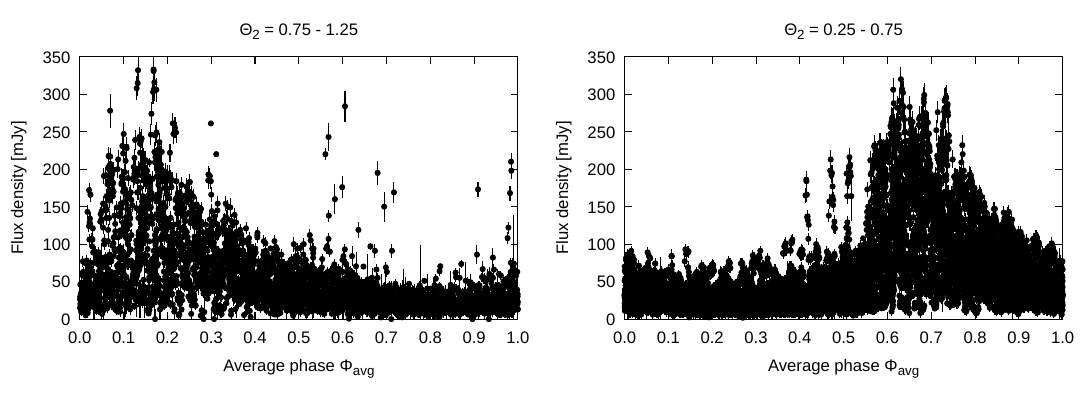}
  \caption{
    Entire radio data set (shown in Fig.~\ref{fig:radioall-lc}) folded over $P_{\rm outburst} = 26.709$~d and separated into double long-term ($2\cdot P_{\rm long}$) phase intervals. For $\Theta_2 = 0.75 - 1.25$, the data cluster around $\Phi_{\rm avg} \approx 0.2$, and for $\Theta_2 = 0.25 - 0.75$, the data cluster around $\Phi_{\rm avg} \approx 0.7$. This is expected from the beating between $P_1$ and $P_2$, as explained in Sect.~\ref{sec:beating}.
  }
  \label{fig:foldPavg}
\end{figure*}

In simple terms, the phenomenon of a beating can be understood by considering the sum of two sine functions oscillating at circular frequencies $\omega_1$ and $\omega_2$,
\begin{equation} \label{eq:beat}
  \sin\omega_1t + \sin\omega_2t =   \sin\underbrace{\frac{\omega_1 + \omega_2}{2}}_{\omega_{\rm avg}}t\cdot \cos\underbrace{\frac{\omega_1 - \omega_2}{2}}_{\omega_{\rm long}/2}t,
\end{equation}
where $\omega_i = 2\pi/P_i$. For a simple demonstration of the concept of beating, we omit any phase offsets (that these are in fact relevant in the multiwavelength context is discussed in \citealt{Jaron2021}). The right-hand side of this equation shows that the sum of the two sine functions can be rewritten as the product of a sine function oscillating at a circular frequency $(\omega_1 + \omega_2)/2$, which is the average of $\omega_1$ and $\omega_2$, and a cosine term that oscillates at $(\omega_1 - \omega_2)/2$. If the difference between $\omega_1$ and $\omega_2$ is small, then the interference pattern has the form of a beating (i.e., a periodic pattern that is slowly modulated). The reason why we define $\omega_{\rm long} := \omega_1 - \omega_2$ as twice the frequency of the cosine term in Eq.~(\ref{eq:beat}) is that what is observed as the long-term modulation in a beating is the envelope of the interference pattern, which oscillates at twice the frequency of the actual cosine term. The reason for this is that the cosine term has a sign flip every half period. When inspecting the interference pattern in detail, one can see that the cosine term indeed has a sign flip during every minimum of the long-term modulation pattern, as we show later in this section.

In the following we show how all of this applies to the radio emission from \lsi{}. The most obvious effect of a beating is the long-term amplitude modulation. This effect is most evident in the GBI data because these data are very well sampled, with several observations per day over a time span of 6.7~years (i.e., 1.5 cycles of the long-term modulation). Analysis of this data set led to the most precise measurement of the long-term period by \citet{Gregory2002}. In the work we present here, we have analyzed the OVRO radio light curve, now spanning three cycles of the long-term modulation, and we detect a very clear signal at the long-term period in the periodogram of Fig.~\ref{fig:ovro-sc}. That the long-term modulation is a decades-spanning stable periodic signal is demonstrated by the autocorrelation study presented in Fig.~\ref{fig:radioall-dcf}.

However, amplitude modulation is not the only effect that is expected from a beating. The other effect is that the individual radio outbursts do not occur with either the orbital or the precession frequency but with a frequency that is the average of the two,
\begin{equation}
  \nu_{\rm outburst} = \frac{\nu_1 + \nu_2}{2} = \frac{1}{2}\left(\frac{1}{P_1} + \frac{1}{P_2}\right) = \frac{1}{P_{\rm outburst}},
\end{equation}
where $\nu_i = P_i^{-1}$. Inserting the most precise measurements of $P_1$ and $P_2$ (i.e., $P_1 = 26.4960 \pm 0.0028$~d \citep{Gregory2002} and $P_2 = 26.926 \pm 0.005$~d \citep{Wu2018}), this results in an expected period of $P_{\rm outburst} = 26.709 \pm 0.003$~d. That the radio outbursts of \lsi{} indeed occur with that period has first been discussed by \citet{Ray1997}, who reported a period of $26.69 \pm 0.02$~d. Furthermore, the slowly oscillating cosine term on the right-hand side of Eq.~(\ref{eq:beat}) has zero crossings that occur with a frequency of $\omega_1 - \omega_2$ (i.e., with a period of $P_{\rm long}$). The effect of this is that the phase of the radio outburst with respect to $P_{\rm outburst}$ performs a phase jump of 0.5 every $P_{\rm long}$ during the minima of the long-term modulation. This has already been pointed out by \citet{Jaron2013}, who demonstrated how this behavior can be used to predict the occurrence of the radio outbursts. In Fig.~\ref{fig:foldPavg}, we show the entire radio archive (i.e., all the data points of the long-term light curve shown in Fig~\ref{fig:radioall-lc}) plotted against the phase of the period $P_{\rm outburst} = 26.709$\,d. In addition, we divide the data into phase intervals with respect to
\begin{equation}
  \Theta_2 = \frac{t - t_0}{2P_{\rm long}} - \mathrm{int}\left(\frac{t - t_0}{2P_{\rm long}}\right),
\end{equation}
which is the phase corresponding to twice the long-term period (i.e., the true period of the cosine term in Eq.~(\ref{eq:beat})). The left panel of Fig.~\ref{fig:foldPavg} shows the data from the interval~$\Theta_2 = 0.75 - 1.25$, and the right panel shows data from the complimentary interval $\Theta_2 = 0.25 - 0.75$. There is a clear separation between the peak occurrence of these two intervals. The data from the former interval have a peak at a position of $\Phi_{\rm avg} \approx 0.2$, and the data from the latter interval peak at $\Phi_{\rm avg} \approx 0.7$. This shows that the radio emission from \lsi{} over the past 45.5~years has been subject to the 0.5 phase jumps in alternating long-term cycles. This is exactly the behavior that is expected from a beating between the $P_1$ and $P_2$ \citep[cf.][]{Jaron2013}.

\subsection{Systematic flux modulation on timescales longer than $P_{\rm long}$?}

A by eye inspection of the full radio light curve shown in Fig.~\ref{fig:radioall-lc} revealed that the peak flux densities reached during each cycle of the long-term modulation are not the same. This is not only the result from the frequency dependency of the flux density \citep[see, e.g.,][]{Massi2009}, which becomes evident when comparing observations carried out at the same or similar radio frequencies. There seems to be a systematic trend with time. In particular, considering the OVRO light curve at 15~GHz observed from MJD~54908 until the end of the time series (see also Fig.~\ref{fig:ovro-lc}) suggests a decline in the amplitude of long-term maxima until about $\Theta \approx 9$. Considering the whole radio data set, this impression seems to fit into the big picture. The highest flux densities are reached during the long-term cycle around $\Theta = 4$ (or $\Theta = 5$, if one wants to put importance on the few data points at $\sim 350$~mJy). After that, there is a systematic trend of decreasing amplitude until the cycle at around $\Theta = 9$. The next long-term cycle has not been fully observed yet, but the flux densities toward $\Theta = 10$ are again higher, hinting at a reversal of the trend (i.e., toward higher amplitudes). Continued radio monitoring of the source will tell whether this impression is confirmed. A caveat to this interpretation of the radio light curve is of course the missing data between $\Theta \approx 5 - 7$. The data points observed at $\Theta \approx 6$ reaching somewhat lower values do not necessarily contradict the hypothesis of a steadily decreasing trend because these data only cover a fraction of one long-term cycle and most probably do not include the maximum of that cycle. Unfortunately, the data base before $\Theta \approx 3.5$ (MJD~49200) is not as good as after that point in time, with the data being considerably sparser. Looking at the data by eye, one might have the impression that there is a systematically increasing trend that starts at a minimum at around $\Theta \approx 1$ (MJD~45000) and lasts until a maximum at $\Theta = 5$, after which the amplitudes follow a decreasing trend. Considering the separation of the flux minima at $\sim 45000$~MJD and $\sim 59500$~MJD would imply that this putative systematic modulation has a characteristic timescale, or probably even period, of $\sim 40$~years.

We conclude these considerations by stating that we see evidence for a variability in the amplitude of long-term radio maxima in the radio emission from \lsi{}, with an indication for a systematic modulation with time. There seem to be hints that this pattern is repeating, but with the current database at hand, we cannot make a firm statement as to whether this is another periodicity of the source, which might be called a "super-long-term modulation."

\section{Conclusions} \label{sec:conclusions}

In this article we have analyzed the updated database of OVRO radio observations of the \gammaray{}-loud X-ray binary~\lsi{} at 15~GHz. In addition, we have combined these data with decades-spanning radio observations of this source in order to investigate the long-term stability of periodic signals in the emission. We conclude the following:
\begin{enumerate}
\item{
  We measure the orbital period as $P_1 = 26.49 \pm 0.05$~d from analysis of the full OVRO data set alone. Analysis of the entire radio archive gives $P_1 = 26.50 \pm 0.12$~d. The larger uncertainty is most likely the result of combining different radio frequencies.
}
\item{
  The presence of a period at $P_2 = 26.93 \pm 0.05$~d is confirmed from the OVRO data alone, and analysis of the full radio archive results in $P_2 = 26.91 \pm 0.11$~d. By combining different frequencies, we obtained a larger uncertainty for the full data set as well. 
}
\item{
    By analyzing each long-term cycle of the OVRO data separately, we showed that the two periods $P_1$ and $P_2$ are present in each of the three consecutive long-term cycles and that their values remain stable.
}
\item{
  The period of the long-term modulation is $P_{\rm long} = 1698 \pm 196$~d in the OVRO data, and it is  $P_{\rm long} = 1601 \pm 152$~d in the full radio data. 
}
\item{
  Analysis of the updated long-term radio archive of \lsi{} using the auto-correlation function revealed that the long-term modulation has now completed ten full cycles with remarkable stability.
}
\end{enumerate}

The presence and the stability of the two close periods, $P_1$ and $P_2$, and the agreement of $P_{\rm long}$ with the beat period of the two combined with the stability of the long-term modulation over ten full cycles strongly favors the interpretation that $P_2$ is the precession period of a relativistic jet. This model has already been successful in reproducing the radio \citep{Massi2014} and gigaelectron volt emission \citep{Jaron2016} from this source. The interpretation of the long-term modulation being the result of changes in the Be star's decretion disk is unlikely because these variations are not expected to remain so stable over such a long period of time, and this has never been observed \citep{Rivinius2013, Massi2016}. 

From inspection of the full radio archive spanning almost 46~years, we obtained the strong impression that there is a systematic modulation of the amplitude of the long-term maxima. However, this data set is still not long enough to make a firm statement. Only the availability of more radio data will allow us to find out if there is another signal in the radio emission from this source. Knowledge about this may improve our understanding of the physical processes at work in this intriguing source. Therefore, continued radio monitoring of \lsi{} is strongly encouraged.

\begin{acknowledgements}
  We thank Gunther Witzel for carefully reading the manuscript and for providing very useful feedback.
  We thank Maria Massi for reading the manuscript and for valuable input during very useful discussions.
  This research has made use of data from the OVRO 40-m monitoring program \citep{Richards2011}, supported by private funding from the California Institute of Technology and the Max Planck Institute for Radio Astronomy, and by NASA grants NNX08AW31G, NNX11A043G, and NNX14AQ89G and NSF grants AST-0808050 and AST- 1109911.
  FJ acknowledges funding by the Austrian Science Fund (FWF) [P35920].
  SK acknowledges support from the European Research Council (ERC) under the European Union's Horizon 2020 research and innovation programme under grant agreement No.~771282. 
\end{acknowledgements}

\bibliographystyle{aa}
\bibliography{jaron2024}

\begin{appendix}

  \section{OVRO monitoring sampling rate} \label{app:sampling}

  \begin{figure}
    \includegraphics{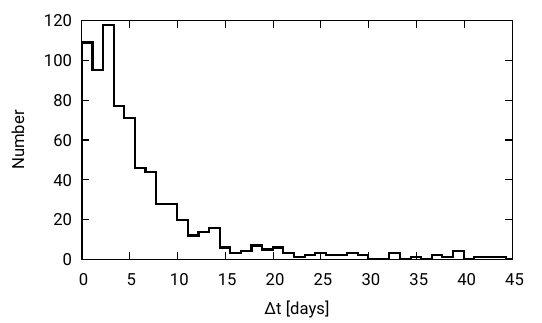}
    \caption{
      Histogram of the time gaps between two adjacent observations within the OVRO data. We only show values until 45~days. There are five longer gaps, the longest being 110~days.
    }
    \label{fig:sampling}
  \end{figure}

  Figure~\ref{fig:sampling} shows a histogram of the time gaps $\Delta t$ between adjacent observations within the OVRO monitoring. We only show values until 45~days for better visual display. There are five larger gaps extending until 110~days, which means that gaps of that length are an exception. 

  \section{Lomb-Scargle false alarm probability} \label{app:fap}

  \begin{figure}
    \includegraphics{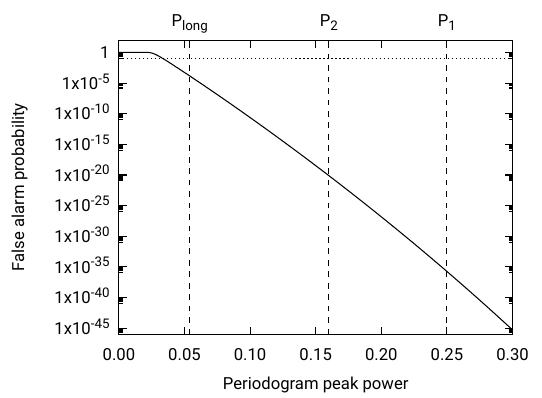}
    \caption{
      False alarm probability plotted against Lomb-Scargle peak power. The vertical dashed lines mark the peak powers of the three periods found in the periodogram. The dotted horizontal line marks a probability of $p = 0.1$. For the data shown here, the \citet{Baluev2008} method has been used as implemented in \texttt{astropy}.  
    }
    \label{fig:fap}
  \end{figure}

  There are four methods for estimating the false alarm probability of periodic features in the Lomb-Scargle implementation in \texttt{astropy}. We tried all four of them, and they gave consistent results. In Fig.~\ref{fig:fap}, we show the results for the \citet{Baluev2008} method, plotting false alarm probability against periodogram peak power. The powers of all three periodic features, $P_1$, $P_2$, and $P_{\rm long}$, are shown by the vertical dashed lines. The period $P_{\rm long}$ has a value of $\sim 10^{-4}$, which is still highly significant. The two periods $P_1$ and $P_2$ have values that are practically in agreement with zero. We are aware that the use of the false alarm probability is only valid in the case of white noise. Indeed, the periodograms obtained from the data analyzed in this article feature only the discussed peaks and otherwise show a very flat noise spectrum.

  \section{Fitting the periodogram} \label{app:fit}

  \begin{table}
    \centering
    \caption{
      Parameters resulting from fitting functions to the periodogram of the OVRO data alone. The upper part of the table shows the results of fitting the superposition of two Gaussians to the double-peak feature. The lower part shows the results from fitting the long-term period with a single Gaussian.
    }
    \label{tab:fitscovro}
    \begin{tabular}{lll}
      \hline
      \hline
      Parameter & Value & Uncertainty\\
      \hline
      $A_1$ & 0.25 & 0.01 \\
      $P_1$ [days] & 26.490 & 0.003 \\
      $\sigma_1$ [days] & 0.046 & 0.002 \\
      $A_2$ & 0.16 & 0.01 \\
      $P_2$ [days] & 26.927 & 0.004 \\
      $\sigma_2$ [days] & 0.047 & 0.004 \\
      \hline
      $A_{\rm long}$ & 0.054 & 0.002 \\
      $P_{\rm long}$ [days] & 1698 & 9 \\
      $\sigma_{\rm long}$ [days] & 196 & 9 \\
      \hline
      \hline
    \end{tabular}
  \end{table}

\begin{table}
    \centering
    \caption{
      Same as Table~\ref{tab:fitscovro} but for the periodogram of all radio data.
    }
    \label{tab:fitscall}
    \begin{tabular}{lll}
      \hline
      \hline
      Parameter & Value & Uncertainty\\
      \hline
      $A_1$ & 0.26 & 0.01 \\
      $P_1$ [days] & 26.500 & 0.004 \\
      $\sigma_1$ [days] & 0.119 & 0.005 \\
      $A_2$ & 0.17 & 0.01 \\
      $P_2$ [days] & 26.911 & 0.007 \\
      $\sigma_2$ [days] & 0.110 & 0.007 \\
      \hline
      $A_{\rm long}$ & 0.056 & 0.001 \\
      $P_{\rm long}$ [days] & 1601 & 4 \\
      $\sigma_{\rm long}$ [days] & 152 & 4 \\
      \hline
      \hline
    \end{tabular}
  \end{table}

  In order to quantify the positions of peaks in the periodograms and their widths, we performed least squares fitting of functions to the periodograms. 
  In general, periodogram peaks are best fit with a Gaussian function of the form
  \begin{equation}
    f(P) = A\,\mathrm{e}^{-\frac{1}{2}\frac{(P - P_i)^2}{\sigma^2}},
  \end{equation}
  with amplitude~$A$, center period~$P_i$, and standard deviation $\sigma$. We took the center period~$P_i$ as the period value and its standard deviation~$\sigma$ as its uncertainty in the period values that we report throughout this article.
  We find that the two-peaked profile of the Lomb-Scargle periodogram is best fit with the function
  \begin{equation}
    f(P) = A_1\,\mathrm{e}^{-\frac{1}{2}\frac{(P - P_1)^2}{\sigma_1^2}} + A_2\,\mathrm{e}^{-\frac{1}{2}\frac{(P - P_2)^2}{\sigma_2^2}},
  \end{equation}
  which is the superposition of two Gaussian functions with amplitudes~$A_{1,2}$; standard deviations~$\sigma_{1,2}$; and center periods~$P_{1,2}$. In this way, we also prove that the two-peaked profile is significantly fitted by a function that is intrinsically two peaked. The results are listed in Table~\ref{tab:fitscovro} for the OVRO data alone and in Table~\ref{tab:fitscall} for the periodogram of the entire archive of radio observations.

\section{Folding over a random period} \label{app:foldrand}

\begin{figure}
    \centering
    \includegraphics{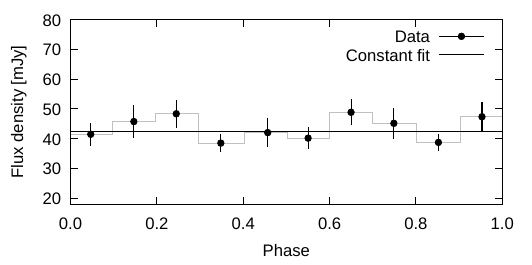}
    \caption{OVRO data folded over the arbitrary period of 30~days. The profile does not show any regular pattern and is reasonably well fitted by a constant, as shown by the solid line.}
    \label{fig:foldPrand}
\end{figure}

In Sect.~\ref{sec:results-fold}, we folded the OVRO data over the periods found by the timing analysis. The results of this are shown in Fig.~\ref{fig:ovro-fold}. For comparison, we folded the data over an arbitrary period of 30~days, shown in Fig.~\ref{fig:foldPrand}. Compared to the data folded over the period peak values, the data folded over the arbitrary period does not show any distinct profile. Figures~\ref{fig:ovro-fold} and \ref{fig:foldPrand} share the same scale of the $y$-axis, which makes it obvious that the variability pattern of the arbitrarily folded data is very flat and does not deviate significantly from the fitted constant, which appears as the solid horizontal line in the plot.
    
\end{appendix}

\end{document}